\documentclass[12pt]{article}
\usepackage{enumerate}
\usepackage{amsfonts}
\usepackage{amsmath}
\usepackage{amssymb}
\usepackage{amsthm}

\def\H{\mathcal{H}}

\def\S{\mathfrak{S}}

\def\T{\mathfrak{T}}

\newcommand{\id}{\mathrm{Id}}
\newcommand{\Tr}{\mathrm{Tr}}

\newcommand{\shs}{\hspace{1pt}}

\newcounter{defin}  \newcounter{lemma}  \newcounter{theorem}
\newcounter{property} \newcounter{corol}  \newcounter{remark} \newcounter{example}

\newenvironment{lemma}{\par\refstepcounter{lemma}
     \textbf{Lemma \thelemma.} }{\rm\par}

\newenvironment{property}{\par\refstepcounter{property}
     \textbf{Proposition \theproperty.}\ }{\rm\par}
\newenvironment{corollary}{\par\refstepcounter{corol}
     \textbf{Corollary \thecorol.} }{\rm\par}
\newenvironment{definition}{\par\refstepcounter{defin}
     \textbf{Definition \thedefin.}\ }{\rm\par}
\newenvironment{remark}{\par\refstepcounter{remark}
     \textbf{Remark \theremark.}}{\rm\par}
\newenvironment{example}{\par\refstepcounter{example}
     \textbf{Example \theexample.}}{\rm\par}

\begin{document}

\title{Upper bounds for the Holevo quantity and their use}
\author{M.E. Shirokov\footnote{Steklov Mathematical Institute, RAS, Moscow, email:msh@mi.ras.ru}}
\date{}
\maketitle

\begin{abstract}
We present a family of easily computable upper bounds for the Holevo quantity of ensemble of quantum states depending on a reference state as a free parameter. These upper bounds are obtained by combining probabilistic and metric characteristics of the ensemble. We show that appropriate choice of the reference state gives tight upper bounds for the Holevo
quantity which in many cases improve existing estimates in the literature.

We also present upper bound for the Holevo quantity of a generalized ensemble of quantum states with finite average energy depending on metric divergence of the ensemble. The specification of this upper bound for the multi-mode quantum oscillator is tight for large energy.

The above results are used to obtain tight upper bounds for the Holevo capacity of  finite-dimensional  and infinite-dimensional energy-constrained quantum channels depending on metric characteristics of the channel output.
\end{abstract}


\pagebreak

\section{Introduction and preliminaries}

The Holevo quantity of ensemble of quantum states (also called Holevo information) is the upper bound for the classical information obtained from  quantum measurements over the ensemble \cite{H-73}. It plays a basic role in analysis of information properties of quantum systems and channels \cite{H-SCI,N&Ch,Wilde}.

The Holevo quantity of a discrete (finite or countable) ensemble $\{p_i,\rho_i\}$ of  quantum states is defined as
$$
\chi\left(\{p_i,\rho_i\}\right)\doteq \sum_{i}p_i H(\rho_i\|\shs\bar{\rho})=H(\bar{\rho})-\sum_{i} p_i H(\rho_i),\quad \bar{\rho}=\sum_{i} p_i\rho_i,
$$
where $H(\cdot\|\cdot)$ is the quantum relative entropy, $H(\cdot)$ is the von Neumann entropy (introduced below) and the second formula is valid if $\,H(\rho_i)<+\infty\,$ for all $i$. So,  the exact value of the Holevo quantity can be found by calculation of the entropy (relative entropy) for a collection of quantum states, which requires some efforts, especially, in the infinite-dimensional case. Therefore it is useful  to have easily computable estimates for the Holevo quantity.

A problem of finding easily computable estimates (in particular, upper estimates) for the Holevo quantity was considered by several authors \cite{Aud,B&H,Roga, Roga+,ZWF}. The main idea of works in this direction is to use  geometrical and probabilistic features of the ensemble to obtain effective estimates. For example, it is shown in \cite{Roga} that in finite dimensions the Holevo quantity is upper bounded by the entropy of the matrices with entries depending on mutual fidelities of states of the ensemble and their probabilities. Recently
Audenaert obtained in \cite{Aud} the following  upper bound:
\begin{equation}\label{A-ub}
\chi(\{p_i,\rho_i\})\leq \upsilon_\mathrm{m}S(\{p_i\}),
\end{equation}
where $\upsilon_\mathrm{m}=\frac{1}{2}\sup_{i,j}\|\rho_i-\rho_j\|_1$ is the maximal trace norm distance between the states of the ensemble and $S(\{p_i\})$ is the Shannon entropy of the probability distribution $\{p_i\}$. It implies that
\begin{equation}\label{A-ub+}
\chi(\{p_i,\rho_i\})\leq \upsilon_\mathrm{m} \log n,
\end{equation}
where $n$ is the number of states in the ensemble $\{p_i,\rho_i\}$.\footnote{In the case $\,n=2\,$ inequality (\ref{A-ub+}) is originally proved in \cite{B&H}.}

Audenaert's upper bound (\ref{A-ub}) refines the well-known rough estimate\break $\chi(\{p_i,\rho_i\})\leq S(\{p_i\})$ by taking metric relations between states of the ensemble into account.

In this paper we present a family of upper bounds for the Holevo quantity depending on a reference state as a free parameter. These upper bounds are proved by applying the Alicki-Fannes-Winter technique (generally used for proving uniform continuity bounds) \cite{A&F,W-CB}. In particular, we obtain several modifications of Audenaert's upper bound (\ref{A-ub}) and of  its corollary (\ref{A-ub+}). We show that the maximal distance $\upsilon_\mathrm{m}$ between states of the ensemble in (\ref{A-ub}) and in (\ref{A-ub+}) can be replaced, respectively, by the quantities
\begin{equation*}
  \varepsilon_{\mathrm{m}}=\textstyle\frac{1}{2}\displaystyle\inf_{\sigma}\sup_i\|\rho_i-\sigma\|_1\quad\textrm{and}\quad \varepsilon_{\mathrm{av}}=\textstyle\frac{1}{2}\displaystyle\inf_{\sigma}\sum_i p_i\|\rho_i-\sigma\|_1
\end{equation*}
called \emph{maximal metric divergence} and \emph{average metric divergence} of the ensemble $\{p_i,\rho_i\}$, which can be significantly less than $\upsilon_\mathrm{m}$.  The cost of such replacement is the appearance of (nonavidable) additional term independent of the size of the ensemble and of the dimension of underlying Hilbert space (Corollaries \ref{amd-ub} and \ref{mmd-ub}).

In the last part of the paper the above results are used to obtain upper bound for the Holevo capacity of a finite-dimensional quantum channel depending on the Chebyshev raduis of its output set. This upper bound gives relatively sharp estimates of the Holevo capacity for several types of channels (in particular, for depolarising and erasure channels). \smallskip

We also present upper bound for the Holevo quantity of a generalized ensemble of quantum states with finite average energy depending on metric divergence of the ensemble and consider its specification for the multi-mode quantum oscillator. This upper bound is used to obtain upper bound for the Holevo capacity of infinite-dimensional quantum channels with energy constraints.\medskip

Let $\mathcal{H}$ be a finite-dimensional or separable infinite-dimensional Hilbert space,
$\mathfrak{B}(\mathcal{H})$ the algebra of all bounded operators with the operator norm $\|\cdot\|$ and $\mathfrak{T}( \mathcal{H})$ the
Banach space of all trace-class
operators in $\mathcal{H}$  with the trace norm $\|\!\cdot\!\|_1$. Let
$\mathfrak{S}(\mathcal{H})$ be  the set of quantum states (positive operators
in $\mathfrak{T}(\mathcal{H})$ with unit trace) \cite{H-SCI,N&Ch,Wilde}.

We denote by $I_{\mathcal{H}}$ the unit operator in a Hilbert space
$\mathcal{H}$ and by $\id_{\mathcal{\H}}$ the identity
transformation of the Banach space $\mathfrak{T}(\mathcal{H})$.\smallskip

A finite or
countable collection $\{\rho_{i}\}$ of states
with a probability distribution $\{p_{i}\}$ is conventionally called
\textit{(discrete) ensemble} and denoted $\{p_{i},\rho_{i}\}$. The state
$\bar{\rho}\doteq\sum_{i}p_{i}\rho_{i}$ is called the \emph{average state} of this  ensemble. \smallskip

The \emph{Shannon entropy} $\,S(\{p_{i}\})=\sum_{i}\eta(p_{i})\,$ of a probability distribution $\{p_i\}$ and the \emph{von Neumann entropy} $\,H(\rho)=\mathrm{Tr}\eta(\rho)\,$ of a
state $\rho\in\mathfrak{S}(\mathcal{H})$, where $\eta(x)=-x\log x$,
have concave homogeneous\footnote{A function $f(x)$ is called homogeneous (of degree 1) if $f(cx)=cf(x)$ for $c\geq0$.}   extensions to the positive cones in $\ell_1$ and in
$\mathfrak{T}(\mathcal{H})$ defined, respectively, by the formulas (cf.\cite{L-2})
\begin{equation}\label{H-ext}
S(\{p_{i}\})=\textstyle\sum_{i}\eta(p_{i})-\eta\!\left(\textstyle\sum_{i}p_{i}\right)\quad \textrm{and} \quad
H(\rho)=\mathrm{Tr}\eta(\rho)-\eta(\mathrm{Tr}\rho).
\end{equation}
The extended von Neumann entropy satisfies the following inequality
\begin{equation}\label{H-ineq}
\textstyle\sum_{i}H(\rho_{i})\leq H\!\left(\textstyle\sum_{i}\rho_{i}\right)\leq
\textstyle\sum_{i}H(\rho_{i})+S\left(\left\{\mathrm{Tr}\rho_{i}\right\}\right),
\end{equation}
valid for any finite or countable collection  $\{\rho_{i}\}$ of positive operators in
$\mathfrak{T}(\mathcal{H})$ with finite
$\sum_i\mathrm{Tr}\rho_{i}$ \cite{N&Ch, O&P}. Denote by $h_2(p)$ the binary entropy $S(\{p,1-p\})$.

\smallskip

The \emph{quantum relative entropy} for two states $\rho$ and
$\sigma$ in $\mathfrak{S}(\mathcal{H})$ is defined as follows
$$
H(\rho\shs\|\shs\sigma)=\sum_i\langle
i|\,\rho\log\rho-\rho\log\sigma\,|i\rangle,
$$
where $\{|i\rangle\}$ is the orthonormal basis of
eigenvectors of the state $\rho$ and it is assumed that
$H(\rho\shs\|\shs\sigma)=+\infty$ if $\,\mathrm{supp}\rho\shs$ is not
contained in $\shs\mathrm{supp}\shs\sigma$ \cite{L-2,O&P}. \smallskip

We will use Donald's identity
\begin{equation}\label{Donald}
\sum_{i}p_{i}H(\rho_{i}\|\shs\sigma)=\sum_{i}p_{i}H(\rho_{i}\|\shs\bar{\rho})+H(\bar{\rho}\shs\|\shs\sigma)
\end{equation}
valid for arbitrary ensemble $\{p_{i},\rho_{i}\}$ of states with the average state $\bar{\rho}$
and arbitrary state $\sigma$ \cite{Don,O&P}.\smallskip

Throughout the paper we will use the following\smallskip

\begin{definition}\label{tight}
An upper bound $g(x)$ for a nonnegative function $f(x)$ on a set $X$ is called tight if $\;\sup_{x\in X}\frac{f(x)}{g(x)}=1$.
\end{definition}

\pagebreak

\section{Estimates for the Holevo quantity}

\subsection{Discrete ensembles}

For arbitrary given  ensemble  $\{p_i,\rho_i\}$ of $\,n\leq\infty\,$ states in $\S(\H)$ and any state $\sigma\in\S(\H)$ consider two ensembles $\{t_i,\tau^+_i\}$ and $\{t_i,\tau^-_i\}$ of $\,n\,$ states in $\S(\H)$, where
$$
t_i=\frac{p_i\|\rho_i-\sigma\|_1}{\sum_ip_i\|\rho_i-\sigma\|_1}\quad\textrm{ and }\quad \tau^{\pm}_i=2\frac{[\rho_i-\sigma]_{\pm}}{\|\rho_i-\sigma\|_1},\quad i=\overline{1,n}
$$
($[\rho_i-\sigma]_+$ and $[\rho_i-\sigma]_-$ are, respectively, the positive and negative parts of the operator $\rho_i-\sigma$). If $\sigma=\rho_{i_0}$ for some $i_0$ then we assume that both ensembles have no states in the $i_0$-th position.
\smallskip

\begin{property}\label{main} \emph{The Holevo quantities of the above ensembles $\{p_i,\rho_i\}$, $\{t_i,\tau^+_i\}$ and $\{t_i,\tau^-_i\}$ are related by  the inequality
\begin{equation}\label{m-ineq}
 \left|\chi(\{p_i,\rho_i\})-\varepsilon\!\left(\chi(\{t_i,\tau^+_i\})-\chi(\{t_i,\tau^-_i\})\right)\right|\leq g(\varepsilon),
\end{equation}
which implies that
\begin{equation}\label{m-ub}
\chi(\{p_i,\rho_i\})\leq \varepsilon\!\left(\chi(\{t_i,\tau^+_i\})-\chi(\{t_i,\tau^-_i\})\right)+g(\varepsilon)\leq \varepsilon\chi(\{t_i,\tau^+_i\})+g(\varepsilon),
\end{equation}
where $\varepsilon=\frac{1}{2}\sum_i p_i\|\rho_i-\sigma\|_1$ and $\,g(\varepsilon)\!\doteq\!(1+\varepsilon)h_2\!\left(\frac{\varepsilon}{1+\varepsilon}\right)$.} \emph{It follows that \footnote{$S$ and $H$ are the homogeneous extensions of the Shannon entropy and of the von Neumann entropy to the positive cones in $\ell_1$ and in $\T(\H)$ defined by the
formulae in (\ref{H-ext}).}
\begin{equation}\label{m-ub-1}
\chi(\{p_i,\rho_i\})\leq \varepsilon S\!\left(\{t_i\}\right)+g(\varepsilon)=S\!\left(\{\textstyle\frac{1}{2}p_i\|\rho_i-\sigma\|_1\}\right)+g(\varepsilon)
\end{equation}
and
\begin{equation}\label{m-ub-2}
\chi(\{p_i,\rho_i\})\leq \varepsilon H\!\left(\sum_i t_i\tau^+_i\right)+g(\varepsilon)=H\!\left(\sum_i p_i[\rho_i-\sigma]_{+}\right)+g(\varepsilon).
\end{equation}}

\emph{Upper bounds (\ref{m-ineq})-(\ref{m-ub-2}) are tight in the sense of Def.\ref{tight}. For any $\,\varepsilon>0$ there is an ensemble $\{p_i,\rho_i\}$ and a state $\sigma$ such that
$\,\varepsilon=\frac{1}{2}\sum_i p_i\|\rho_i-\sigma\|_1$ and}
$$
\chi(\{p_i,\rho_i\})-\varepsilon\!\left(\chi(\{t_i,\tau^+_i\})-\chi(\{t_i,\tau^-_i\})\right)=h_2(\varepsilon).
$$
\end{property}\smallskip

\begin{remark}\label{r-1}
The last assertion of Proposition \ref{main} shows that the right hand side of (\ref{m-ineq}) can not be less than $h_2(\varepsilon)$, which is equivalent to $g(\varepsilon)$ for small $\varepsilon$.\medskip
\end{remark}

\emph{Proof.} Inequality
(\ref{m-ineq}) directly follows from Proposition 1 in \cite{CHI} (with trivial C). It suffices to take the $qc$-states
$$
\rho_{AB}=\sum_{i=1}^n p_i\rho_i\otimes |i\rangle\langle i|\quad \textrm{and} \quad\sigma_{AB}=\sum_{i=1}^n p_i\sigma\otimes |i\rangle\langle i|,
$$
where $\H_A=\H$ and $\{|i\rangle\}$ is an orthonormal basis in $n$-dimensional Hilbert space $\H_B$, and to note that $\rho_B=\sigma_B$,
$$
I(A\!:\!B)_{\rho}=\chi(\{p_i,\rho_i\}),\quad I(A\!:\!B)_{\sigma}=0\quad\textrm{and}\quad I(A\!:\!B)_{\tau_\pm}=\chi(\{t_i,\tau^\pm_i\}),
$$
where $\tau_\pm=\varepsilon^{-1}[\rho-\sigma]_\pm$. Inequalities (\ref{m-ub-1}) and (\ref{m-ub-2}) directly follow from (\ref{m-ub}).\smallskip

The tightness of upper bounds (\ref{m-ineq})-(\ref{m-ub-2}) and the last assertion of the proposition can be shown by using Examples \ref{one} and \ref{two} below. $\square$

\smallskip

Note first that Proposition \ref{main} implies the following easily computable upper bounds for  the Holevo quantity.\smallskip

\begin{corollary}\label{main-cor} \emph{The Holevo quantity $\chi(\{p_i,\rho_i\})$ of an arbitrary ensemble $\{p_i,\rho_i\}$ of $\,n\leq\infty\,$ states in $\S(\H)$ is upper bounded by any of the quantities
\begin{equation}\label{ec-ub}
\textstyle\frac{1}{2}\sup_i\|\rho_i-\sigma\|_1S(\{p_j\})+g(\varepsilon),\quad \varepsilon\log n+g(\varepsilon),\quad \varepsilon\log d+g(\varepsilon),
\end{equation}
where $\sigma$ is any state in $\,\S(\H)$, $\,\varepsilon=\frac{1}{2}\sum_i p_i\|\rho_i-\sigma\|_1$ and $\,d=\dim\H\leq\infty$.}
\medskip
\end{corollary}

The first and the second upper bound in (\ref{ec-ub}) may be stronger than Audenaert's upper bound (\ref{A-ub}) and its corollary (\ref{A-ub+}) correspondingly (despite nonavoidable term $g(\varepsilon)$ in the formers), since the values of $\sup_i\|\rho_i-\sigma\|_1$ and $\sum_i p_i\|\rho_i-\sigma\|_1$ may be significantly  less than
$\,\sup_{i,j}\|\rho_i-\rho_j\|_1$ for ensembles with arbitrary large Holevo quantity (see Examples \ref{three} and \ref{MMD-e} below).\medskip

Proposition \ref{main} shows that the quantity
$$
T_{\chi}(\{p_i,\rho_i\}|\shs\sigma)\doteq\varepsilon\!\left(\chi(\{t_i,\tau^+_i\})-\chi(\{t_i,\tau^-_i\})\right)
$$
can be considered as an approximation of $\,\chi(\{p_i,\rho_i\})$.

We will call the quantity
\begin{equation}\label{r-amd}
\varepsilon=\frac{1}{2}\sum_i p_i\|\rho_i-\sigma\|_1
\end{equation}
\emph{metric divergence of an ensemble $\{p_i,\rho_i\}$ with respect to a state} $\sigma$ and will denote it by $D(\{p_i,\rho_i\}|\shs\sigma)$.
\smallskip

The \emph{reference state} $\sigma$ is a free parameter which can be used to optimise upper bounds (\ref{m-ineq})-(\ref{ec-ub}). Below we will specify these upper bounds and analyse the quantity $T_{\chi}(\{p_i,\rho_i\}|\shs\sigma)$ in the following cases:
\begin{itemize}
  \item $\sigma=\rho_c\doteq I_{\H}/d\,$ is  the chaotic state in $d$-dimensional Hilbert space $\H$;
  \item $\sigma=\bar{\rho}\doteq \sum_ip_i\rho_i\,$ is  the average state of the ensemble $\{p_i,\rho_i\}$;
  \item $\sigma=\rho_{i_0}\,$ is one of the states of the ensemble $\{p_i,\rho_i\}$;
  \item $\sigma\,$ is the state minimazing the value of $\,\frac{1}{2}\sum_i p_i\|\rho_i-\sigma\|_1$;
  \item $\sigma\,$ is the state minimazing the value of $\,\frac{1}{2}\sup_i\|\rho_i-\sigma\|_1$.
\end{itemize}

\textbf{Note:} The minimazing states $\sigma$ in the last two cases may not coincide with each other and with the average state $\bar{\rho}$  even for ensemble
$\{p_i,\rho_i\}$ of isomorphic states with uniform probability distribution $\{p_i\}$ (see Example 4 below).\smallskip

\textbf{The case $\sigma=\rho_c$.} In this case the values of $\|\rho_i-\sigma\|_1$ and the ensembles $\{t_i,\tau^+_i\}$ and $\{t_i,\tau^-_i\}$ are easily determined. Indeed, if $\rho=\sum_{k}\lambda_k|\varphi_k\rangle\langle\varphi_k|$ is a spectral decomposition of a state $\rho$ in $d$-dimensional Hilbert space $\H$ then
$$
 [\rho-\rho_c]_+=\sum_{\lambda_k>1/d}(\lambda_k-1/d\d)|\varphi_k\rangle\langle\varphi_k|,\quad [\rho-\rho_c]_-=\sum_{\lambda_k<1/d}(1/d-\lambda_k)|\varphi_k\rangle\langle\varphi_k|
$$
and $\|\rho-\rho_c\|_1=\sum_{k}|\lambda_k-1/d|$.  It follows, in particular, that in this case the probability distribution $\{t_i\}$ is completely determined by eigenvalues of the states $\rho_i$ and by the probability distribution $\{p_i\}$.\smallskip

The above formulae show that $\{t_i,\tau^+_i\}=\{p_i,\rho_i\}$ for any ensemble $\{p_i,\rho_i\}$ consisting of states proportional to projectors of the same rank.\smallskip

\begin{example}\label{one} Let $\{p_i,\rho_i\}$ be an arbitrary ensemble of pure states. Then $\frac{1}{2}\|\rho_i-\rho_c\|_1=1-1/d$, $\,t_i=p_i$,
$\,\tau^+_i=\rho_i\,$ and $\,\tau^-_i=\tilde{\rho}_i\doteq(d-1)^{-1}(I_{\H}-\rho_i)$. So,
$$
T_{\chi}(\{p_i,\rho_i\}|\shs\rho_c)=(1-1/d)\left(\chi(\{p_i,\rho_i\})-\chi(\{p_i,\tilde{\rho}_i\})\right)
$$
and hence 
$$
\chi(\{p_i,\rho_i\})-T_{\chi}(\{p_i,\rho_i\}|\shs\rho_c)=(1/d)\chi(\{p_i,\rho_i\})+(1-1/d)\chi(\{p_i,\tilde{\rho}_i\})
$$
Since $\chi(\{p_i,\tilde{\rho}_i\})\leq\log d-\log(d-1)$, we have
$$
0\,\leq\,\chi(\{p_i,\rho_i\})-T_{\chi}(\{p_i,\rho_i\}|\shs\rho_c)\,\leq\,\frac{\log d}{d}+(1-1/d)\log\frac{d}{d-1}=h_2(1/d),
$$
where an equality holds in the second inequality if and only if $\,\bar{\rho}=\rho_c$.  

The upper bounds (\ref{m-ub-1}) and (\ref{m-ub-2}) imply, respectively,
$$
\chi(\{p_i,\rho_i\})\leq(1-1/d)S(\{p_i\})+g(1-1/d)
$$
and
$$
\chi(\{p_i,\rho_i\})\leq(1-1/d)H(\bar{\rho})+g(1-1/d),
$$
where $\bar{\rho}\doteq\sum_ip_i\rho_i$.
We see that the second upper bound is closer to the exact value $H(\bar{\rho})$ of $\chi(\{p_i,\rho_i\})$.
\end{example}\smallskip

\begin{example}\label{two} Let $\{p_i,\rho_i\}$ be an ensembles of states proportional to $k$-rank projectors in $d$-dimensional Hilbert space $\H$ such that $\sum_i p_i\rho_i=\rho_c$. If $\sigma=\rho_c$ then it is easy to see that $\,\varepsilon=\frac{d-k}{d}$,  $\,\{t_i,\tau^+_i\}=\{p_i,\rho_i\}$ and that the ensemble $\,\{t_i,\tau^-_i\}$ consists of states proportional to $(d-k)$-rank projectors and has the average state $\rho_c$. It follows that
$$
\chi(\{p_i,\rho_i\})-T_{\chi}(\{p_i,\rho_i\}|\shs\rho_c)=
\frac{k}{d}\log\frac{d}{k}+\frac{d-k}{d}\log\frac{d}{d-k}=h_2(\varepsilon).
$$
This is the first example proving the last assertion of Proposition \ref{main}.
\end{example}\smallskip

\textbf{The case $\sigma=\bar{\rho}$.} For each $i$ let $\hat{\rho}_i=(1-p_i)^{-1}\sum_{j\neq i}p_j\rho_j$ be the complementary state to the state $\rho_i$ \cite{Aud}. Then $\rho_i-\bar{\rho}=(1-p_i)(\rho_i-\hat{\rho}_i)$. So, in this case
\begin{equation}\label{a-eps-}
\tau^\pm_i=\frac{2[\rho_i-\hat{\rho}_i]_\pm}{\|\rho_i-\hat{\rho}_i\|_1}\quad\textrm{and}\quad t_i=\frac{1}{2\varepsilon}p_i\|\rho_i-\bar{\rho}\|_1=\frac{1}{2\varepsilon}p_i(1-p_i)\|\rho_i-\hat{\rho}_i\|_1
\end{equation}
for $i=\overline{1,n}$, where
\begin{equation}\label{a-eps}
\varepsilon=D(\{p_i,\rho_i\}|\shs\bar{\rho}\shs)\doteq\frac{1}{2}\sum_{i=1}^np_i\|\rho_i-\bar{\rho}\|_1=\frac{1}{2}\sum_{i=1}^np_i(1-p_i)\|\rho_i-\hat{\rho}_i\|_1.
\end{equation}
By convexity of the trace norm we have
\begin{equation}\label{a-eps+}
\|\rho_i-\hat{\rho}_i\|_1\leq \upsilon_\mathrm{m}\quad \textrm{and hence}\quad \varepsilon\leq \upsilon_\mathrm{m}\! \left(1-\sum_{i=1}^n p_i^2\right),
\end{equation}
where $\upsilon_\mathrm{m}=\frac{1}{2}\sup_{i,j}\|\rho_i-\rho_j\|_1$. \smallskip

In the case $\,\sigma=\bar{\rho}\,$ the ensembles $\,\{t_i,\tau^+_i\}$ and $\,\{t_i,\tau^-_i\}$ have the same average state. So, if this average state has finite entropy then
$$
\begin{array}{rl}
\displaystyle T_{\chi}(\{p_i,\rho_i\}|\shs\bar{\rho}\shs)&\displaystyle\!\!=\;\varepsilon\sum_{i=1}^n t_i(H(\tau^-_i)-H(\tau^+_i))\\&\displaystyle\!\!=\;\sum_{i=1}^n p_i(1-p_i)\left(H([\rho_i-\hat{\rho}_i]_-)-H([\rho_i-\hat{\rho}_i]_+)\right).
\end{array}
$$
If the ensemble $\{p_i,\rho_i\}$ consists of mutually orthogonal states then
$$
[\rho_i-\hat{\rho}_i]_+=\rho_i,\quad [\rho_i-\hat{\rho}_i]_-=\hat{\rho}_i, \quad(1-p_i)H(\hat{\rho}_i)=H(\bar{\rho})-p_iH(\rho_i)-h_2(p_i)
$$
and hence
$$
T_{\chi}(\{p_i,\rho_i\}|\shs\bar{\rho}\shs)=\sum_i p_i(1-p_i)\left(H(\hat{\rho}_i)-H(\rho_i)\right)=\chi(\{p_i,\rho_i\})-\sum_i p_ih_2(p_i).
$$
We see again that the quantity  $T_{\chi}$ may be less than the Holevo quantity. Since in this case $\,\varepsilon=1-\sum_i p_i^2$, by the concavity of $h_2$ we have
$$
\chi(\{p_i,\rho_i\})-T_{\chi}(\{p_i,\rho_i\}|\shs\bar{\rho}\shs)=\sum_{i=1}^n p_ih_2(p_i)\leq h_2\!\left(\sum_{i=1}^np^2_i\right)=h_2(\varepsilon)\leq g(\varepsilon)
$$
in accordance with (\ref{m-ub}).\smallskip

By using (\ref{a-eps-})-(\ref{a-eps+}) the upper bounds in Proposition \ref{main} and Corollary \ref{main-cor} in the case $\sigma=\bar{\rho}\shs$  can be specified as follows\smallskip
\begin{corollary}\label{a-state} \emph{Let $\,\{p_i,\rho_i\}$ be an ensembles of $\,n\leq+\infty$ states in $\,\S(\H)$ and $\,d=\dim\H\leq+\infty$. Then \footnote{$S$ is the homogenious enstensions of the Shannon entropy to the positive cone in $\ell_1$ defined by the first
formula in (\ref{H-ext}).}
\begin{equation}\label{a-state-1}
\!\begin{array}{rl}
\chi(\{p_i,\rho_i\})\leq & \!\!\!S(\{\frac{1}{2}p_i(1-p_i)\|\rho_i-\hat{\rho}_i\|_1\})+g(\varepsilon)\\\\\leq & \!\!\!\upsilon_\mathrm{m} S(\{p_i(1-p_i)\})+g(\varepsilon)
\leq \upsilon_\mathrm{m} (1-\sum_i p_i^2) \log n+g(\varepsilon)
\end{array}
\end{equation}
and
\begin{equation}\label{a-state-2}
\textstyle\chi(\{p_i,\rho_i\})\leq \varepsilon \log d+g(\varepsilon)\leq \upsilon_\mathrm{m} (1-\sum_i p_i^2) \log d+g(\upsilon_\mathrm{m}(1-\sum_i p_i^2)),
\end{equation}
where $\,\varepsilon=D(\{p_i,\rho_i\}|\shs\bar{\rho}\shs)$ determined in (\ref{a-eps}) and $\,\upsilon_\mathrm{m}=\frac{1}{2}\sup_{i,j}\|\rho_i-\rho_j\|_1$.
The term $g(\varepsilon)$ in all the inequalities in (\ref{a-state-1}) can be replaced by $\,g(\upsilon_\mathrm{m}(1-\sum_i p_i^2))$.}
\end{corollary}\smallskip

The last upper bound in (\ref{a-state-1}) is stronger than (\ref{A-ub+}) for ensembles with significantly non-uniform probability distribution (for which $1-\sum_i p_i^2\ll 1$).\smallskip

\begin{example}\label{three} Let $\{p_i,\rho_i\}$ be an ensembles of $\,n+1\,$ mutually orthogonal states, where $p_1=1-\delta$ and $p_i=\delta/n$ for $i=\overline{2,n+1}$. Then $\upsilon_\mathrm{m}=1$ and $1-\sum_i p_i^2=2\delta-(1+1/n)\delta^2$. So, the last upper bound in (\ref{a-state-1})  gives
$$
\chi(\{p_i,\rho_i\})\leq (2\delta-(1+1/n)\delta^2)\log n+g(2\delta-(1+1/n)\delta^2),
$$
while $\chi(\{p_i,\rho_i\})=S(\{p_i\})=\delta\log n+h_2(\delta)$.
We see that the term $1-\sum_i p_i^2$ allows to take degeneracy of the probability distribution $\{p_i\}$ into account.\smallskip
\end{example}

\textbf{The case $\sigma=\rho_{i_0}$.} We will assume that $i_0=1$. In this case
\begin{equation}\label{1-eps}
\tau^\pm_i=\frac{2[\rho_i-\rho_1]_\pm}{\|\rho_i-\rho_1\|_1},\quad t_i=\frac{1}{2\varepsilon}p_i\|\rho_i-\rho_1\|_1, \quad i=\overline{2,n},
\end{equation}
where
\begin{equation}\label{1-eps+}
\varepsilon=D(\{p_i,\rho_i\}|\shs\rho_1)\doteq\frac{1}{2}\sum_{i=2}^n p_i\|\rho_i-\rho_1\|_1\leq 1-p_1.
\end{equation}
If the state $\rho_1$ is orthogonal to all other states of the ensemble then $\varepsilon=1-p_1$ and
$$
\tau^+_i=\rho_i,\quad\tau^-_i=\rho_1,\quad t_i=\tilde{p}^1_i\doteq p_i(1-p_1)^{-1}\quad i=\overline{2,n}.
$$
So, in this case $\,\chi(\{t_i,\tau^+_i\})=\chi(\{\tilde{p}^1_i,\rho_i\}_{i>1})$ and  $\,\chi(\{t_i,\tau^-_i\})=0$. Hence
$$
T_{\chi}(\{p_i,\rho_i\}|\shs\rho_1)=(1-p_1)\chi(\{\tilde{p}^1_i,\rho_i\}_{i>1}),
$$
while Donald's identity (\ref{Donald}) implies  that
$$
\chi(\{p_i,\rho_i\})=(1-p_1)\chi(\{\tilde{p}^1_i,\rho_i\}_{i>1})+h_2(1-p_1).
$$
This is the second example proving the last assertion of Proposition \ref{main}. \medskip

By using (\ref{1-eps})-(\ref{1-eps+}) and the equality $\,S(\{p_i\}_{i\geq0})=S(\{p_i\}_{i>0})+h_2(p_1)\,$ the upper bounds in Proposition \ref{main} and Corollary \ref{main-cor} in the case $\sigma=\rho_1$  can be specified as follows\smallskip
\begin{corollary}\label{1-state} \emph{Let $\,\{p_i,\rho_i\}$ be an ensembles of $\,n\leq+\infty$ states in $\,\S(\H)$ and $\,d=\dim\H\leq+\infty$. Then
\begin{equation}\label{1-state-1}
\begin{array}{rl}
\chi(\{p_i,\rho_i\})\leq &\!\!\! \varepsilon_1 S\left(\{\frac{1}{2\varepsilon_1}p_i\|\rho_i-\rho_1\|_1\}_{i>1}\right)+g(\varepsilon_1)\\\\
\leq & \!\!\! \upsilon_1S(\{p_i\})+[g((1-p_1)\upsilon_1)-\upsilon_1h_2(1-p_1)]
\end{array}
\end{equation}
and
\begin{equation}\label{1-state-2}
\chi(\{p_i,\rho_i\})\leq \varepsilon_1\log d+g(\varepsilon_1)\leq \upsilon_1(1-p_1)\log d+g(\upsilon_1(1-p_1)),
\end{equation}
where $\,\varepsilon_1=\frac{1}{2}\sum_{i>1}p_i\|\rho_i-\rho_1\|_1\,$ and $\,\upsilon_1=\frac{1}{2}\sup_{i>1}\|\rho_i-\rho_1\|_1$.}
\end{corollary}\medskip

Upper bounds in (\ref{1-state-1}) are modifications of Audenaert's upper bound (\ref{A-ub}). The term in square brackets in the second of them is
equal to
$$
\upsilon_1(1-p_1)(-\log \upsilon_1)+o\shs(1-p_1)
$$
for $p_1$ close to $1$. This term is the cost for replacing the maximal distance $\upsilon_\mathrm{m}$ between all states of ensemble in (\ref{A-ub}) by the maximal distance $\upsilon_1$ from the first state of ensemble to all others. It is easy to find an ensemble $\{p_i,\rho_i\}$ with arbitrary $S(\{p_i\})$  such that $\upsilon_1$ is significantly less than $\upsilon_\mathrm{m}$  (such ensemble can be obtained by adding the state $|1\rangle\langle 1|$ to the ensemble in Example \ref{MMD-e} below).\medskip

\textbf{The average metric divergence.} For a given ensemble $\{p_i,\rho_i\}$ consider the quantity
\begin{equation}\label{amd}
  \varepsilon_{\mathrm{av}}(\{p_i,\rho_i\})=\textstyle\frac{1}{2}\displaystyle\inf_{\sigma}\sum_i p_i\|\rho_i-\sigma\|_1,
\end{equation}
which can be called \emph{average metric divergence} of the ensemble $\{p_i,\rho_i\}$. In finite dimensions
the infimum in (\ref{amd}) is always achieved at some state $\sigma$ which
will be called \emph{AMD-optimal state} for the ensemble $\{p_i,\rho_i\}$. For the ensemble of two states $\rho_1$ and $\rho_2$ with probabilities $p_1$ and $p_2=1-p_1$ AMD\nobreakdash-\hspace{0pt}optimal states  are easily determined: if $p_1>p_2$ (correspondingly, $p_1<p_2$) then $\rho_2$ (correspondingly, $\rho_1$) is a unique
AMD-optimal state, if $p_1=p_2$ then any convex mixture of the states $\rho_1$ and $\rho_2$ is an AMD-optimal state for this ensemble. In this case $\varepsilon_{\mathrm{av}}=\frac{1}{2}\min\{p_1,p_2\}\|\rho_1-\rho_2\|_1$. In general,  continuity and convexity of the function  $\sigma\mapsto\sum_i p_i\|\rho_i-\sigma\|_1$   implies that the set of all AMD-optimal states for a given ensemble is closed and convex. The below example shows (contrary to  intuition) that the average state $\bar{\rho}$ of an ensemble of isomorphic states with uniform probability distribution may be not AMD-optimal.
\smallskip

\begin{example}\label{AMD-opt} Let $\{p_i,|\varphi_i\rangle\langle\varphi_i|\}_{i=1}^4$ be the ensemble of four pure states in $\mathrm{3\textup{-}D}$ Hilbert space $\H$, where $p_i\equiv 1/4,$  $|\varphi_1\rangle=|1\rangle$, $|\varphi_2\rangle=-\frac{1}{2}|1\rangle+\frac{\sqrt{3}}{2}|2\rangle$,  $|\varphi_3\rangle=-\frac{1}{2}|1\rangle-\frac{\sqrt{3}}{2}|2\rangle$ and $|\varphi_4\rangle=|3\rangle$ (here $\{|1\rangle,|2\rangle,|3\rangle\}$ is an orthonormal basis in $\H$). Then $\bar{\rho}=\frac{3}{8}(|1\rangle\langle 1|+|2\rangle\langle 2|)+\frac{1}{4}|3\rangle\langle 3|$. It is easy to see that
$$
\frac{1}{2}\sum_{i=1}^4 p_i\||\varphi_i\rangle\langle\varphi_i|-\bar{\rho}\|_1=\frac{21}{32}>\frac{5}{8}=\frac{1}{2}\sum_{i=1}^4 p_i\||\varphi_i\rangle\langle\varphi_i|-\sigma\|_1,
$$
where $\,\sigma=\frac{1}{2}(|1\rangle\langle 1|+|2\rangle\langle 2|)$ is a unique AMD-optimal state for this ensemble.
\end{example}
\medskip

By taking AMD-optimal state\footnote{If $\dim\H=+\infty$ and there are no AMD-optimal states, it suffices to take for given $\epsilon>0$ a state $\sigma_\epsilon$ such that $\,\frac{1}{2}\sum_i p_i\|\rho_i-\sigma_\epsilon\|_1$ is $\epsilon$-close to $\varepsilon_{\mathrm{av}}$.} in the role of the reference state $\sigma$ in Corollary \ref{main-cor} we obtain the following \smallskip
\begin{corollary}\label{amd-ub} \emph{Let $\,\{p_i,\rho_i\}$ be an ensembles of $\,n\leq+\infty$ states in $\,\S(\H)$ and $\,d=\dim\H\leq+\infty$. Then
$$
\chi(\{p_i,\rho_i\})\leq \varepsilon_{\mathrm{av}}\log n+g(\varepsilon_{\mathrm{av}})\quad\textrm{and}\quad\chi(\{p_i,\rho_i\})\leq \varepsilon_{\mathrm{av}}\log d+g(\varepsilon_{\mathrm{av}}),
$$
where $\varepsilon_{\mathrm{av}}$ is the average metric divergence of $\,\{p_i,\rho_i\}$ defined in (\ref{amd}).}
\end{corollary}\smallskip

Since $\varepsilon_{\mathrm{av}}$ may be significantly less than the maximal distance $\upsilon_\mathrm{m}$ between states of an ensemble  $\{p_i,\rho_i\}$, the first upper bound in Corollary \ref{amd-ub} may be stronger than upper bound (\ref{A-ub+}) despite (nonavoidable) additional term $g(\varepsilon_{\mathrm{av}})$.\medskip

\textbf{The maximal  metric divergence.} For a given ensemble $\{p_i,\rho_i\}$ consider the quantity
\begin{equation}\label{mmd}
  \varepsilon_{\mathrm{m}}(\{p_i,\rho_i\})=\textstyle\frac{1}{2}\displaystyle\inf_{\sigma}\sup_i\|\rho_i-\sigma\|_1,
\end{equation}
which can be called \emph{maximal metric divergence} of the ensemble $\{p_i,\rho_i\}$. In finite dimensions
the infimum in (\ref{mmd}) is always achieved at some state $\sigma$ which
will be called \emph{MMD-optimal state} for the ensemble $\{p_i,\rho_i\}$. For ensemble of two states $\rho_1$ and $\rho_2$ with any probabilities $p_1$ and $p_2=1-p_1$ the state $\frac{1}{2}(\rho_1+\rho_2)$ is a unique  MMD-optimal state. In this case $\varepsilon_{\mathrm{m}}=\frac{1}{4}\|\rho_1-\rho_2\|_1$.\smallskip

The ensemble of four pure states in Example \ref{AMD-opt} has a unique MMD\nobreakdash-\hspace{0pt}optimal state $\rho_c\doteq I_{\H}/3$ not coinciding with the average state and with the AMD\nobreakdash-\hspace{0pt}optimal state of this ensemble. For this ensemble $\varepsilon_{\mathrm{m}}=2/3>\varepsilon_{\mathrm{av}}=5/8$.

\smallskip

By taking MMD\nobreakdash-\hspace{0pt}optimal state in the role of the reference state $\sigma$ in Corollary  \ref{main-cor}  we obtain the following\smallskip

\begin{corollary}\label{mmd-ub} \emph{Let $\,\{p_i,\rho_i\}$ be an ensembles of $\,n\leq+\infty$ states in $\,\S(\H)$, where $\,\dim\H\leq+\infty$. Then
$$
\chi(\{p_i,\rho_i\})\leq \varepsilon_{\mathrm{m}}S(\{p_i\})+g(\varepsilon_{\mathrm{m}}),
$$
where $\varepsilon_{\mathrm{m}}$ is the maximal metric divergence of $\{p_i,\rho_i\}$ defined in (\ref{mmd}).}\smallskip
\end{corollary}

We will show that in some cases this upper bound is stronger than the Audenaert's upper bound (\ref{A-ub})
despite (nonavoidable) extra term $g(\varepsilon_{\mathrm{m}})$ (bounded by $g(1)=2\log2$). Note first that
$$
\varepsilon_{\mathrm{m}}\leq\textstyle\frac{1}{2}\sup_i\|\rho_i-\bar{\rho}\|_1\leq \upsilon_{\mathrm{m}}
$$
by convexity of the trace norm.\smallskip

For any ensemble of two states we have $\upsilon_{\mathrm{m}}/\varepsilon_{\mathrm{m}}=2$, but for multi-state ensembles the difference between
$\varepsilon_{\mathrm{m}}$ and $\upsilon_{\mathrm{m}}$ are not so large.\footnote{I would be grateful for any comments concerning possible values of $\upsilon_{\mathrm{m}}/\varepsilon_{\mathrm{m}}$ in general case.} The following example shows existence of ensemble with arbitrary large Holevo quantity for which $\upsilon_{\mathrm{m}}/\varepsilon_{\mathrm{m}}$ is close to $\sqrt{2}$.  \smallskip

\begin{example}\label{MMD-e} Let $\{p_i,|\varphi_i\rangle\langle\varphi_i|\}_{i=1}^n$ be the ensemble of $n$ pure states in $(n+1)\textup{-}$dimensional Hilbert space $\H$, where $\{p_i\}$ is an arbitrary probability distribution and $|\varphi_i\rangle=\sqrt{1-a^2}|1\rangle+a|i+1\rangle$, $a\in[0,1]$ (here $\{|1\rangle,...,|n+1\rangle\}$ is an orthonormal basis in $\H$). Then
$$
\||\varphi_i\rangle\langle\varphi_i|-|\varphi_j\rangle\langle\varphi_j|\|_1=2\sqrt{1-|\langle\varphi_i|\varphi_j\rangle|^2}=2a\sqrt{2-a^2}
$$
and
$$
\||\varphi_i\rangle\langle\varphi_i|-|1\rangle\langle 1|\|_1=2\sqrt{1-|\langle\varphi_i|1\rangle|^2}=2a.
$$
It follows that $\upsilon_{\mathrm{m}}=a\sqrt{2-a^2}$, while $\varepsilon_{\mathrm{m}}\leq a$.\footnote{One can show that $\,n^{-1}\sum_{i=1}^n|\varphi_i\rangle\langle\varphi_i|\,$ is a unique MMD-optimal state for this ensemble  and that $\,\varepsilon_{\mathrm{m}}=a-o(a)<a$.} So, in this case  Audenaert's upper bound (\ref{A-ub}) and the upper bound in Corollary \ref{mmd-ub} give, respectively,
$$
\chi(\{p_i,|\varphi_i\rangle\langle\varphi_i|\})\leq a\sqrt{2-a^2}S(\{p_i\}),
$$
and
$$
\chi(\{p_i,|\varphi_i\rangle\langle\varphi_i|\})\leq aS(\{p_i\})+g(a).
$$
It is clear that the latter upper bound is stronger than the former for small $a$ and large $S(\{p_i\})$.

Direct calculation of eigenvalues of the state $\bar{\rho}\doteq\sum_{i=1}^n p_i|\varphi_i\rangle\langle\varphi_i|\,$ in the case $\,p_i\equiv1/n\,$ shows that
$$
\chi(\{p_i,|\varphi_i\rangle\langle\varphi_i|\})=H(\bar{\rho})=(1-1/n)a^2\log(n-1)+h_2\!\left((1-1/n)a^2\right).\;\;\square
$$
\end{example}

\subsection{Generalized ensembles with finite average energy}

In analysis of infinite-dimensional quantum systems and channels it is necessary to consider \textit{generalized}  ensembles of quantum states defined as  Borel probability measures on the set of quantum states \cite{H-SCI,H-Sh-2}. A discrete ensemble $\{p_i,\rho_i\}$ corresponds to the measure $\sum_i p_i\delta(\rho_i)$, where $\delta(\rho)$ is the Dirac measure concentrating at a state $\rho$.  The average state of a generalized
ensemble $\mu$ is the barycenter of the measure
$\mu $ defined by the Bochner integral
\begin{equation*}
\bar{\rho}(\mu )=\int\rho \mu (d\rho ).
\end{equation*}

The Holevo quantity of a
generalized ensemble $\mu$ is defined as
\begin{equation*}
\chi(\mu)=\int H(\rho\shs \|\shs \bar{\rho}(\mu))\mu (d\rho )=H(\bar{\rho}(\mu))-\int H(\rho)\mu (d\rho ),  
\end{equation*}%
where the second formula is valid under the condition $H(\bar{\rho}(\mu))<+\infty$ \cite{H-SCI, H-Sh-2}.\smallskip

In this subsection we consider upper bounds for the Holevo quantity of generalised ensembles $\,\mu\,$ with finite average energy
$$
\bar{E}(\mu)\doteq \Tr H\bar{\rho}(\mu)=\int\Tr H\rho\,\mu(d\rho),
$$
provided that the Hamiltonian $H$ of the system satisfies the condition
\begin{equation}\label{g-cond}
\Tr e^{-\lambda H}<+\infty\,\textrm{ for some }\,\lambda>0.
\end{equation}
Condition (\ref{g-cond}) implies that all spectral projectors of $H$ corresponding to finite intervals are finite-dimensional and that
the von Neumann entropy $H(\rho)$ is bounded on the sets of states $\rho$ with bounded energy $E(\rho)\doteq\Tr H\rho$ \cite[Pr.1]{EC}. It follows that
\begin{equation}\label{F-def}
F_{H}(E)\doteq\sup_{\Tr H\rho\leq E}H(\rho)
\end{equation}
is a finite function on $[E_0, +\infty)$, where
$E_0\doteq \inf_{\|\varphi\|=1}\langle\varphi|H|\varphi\rangle$.

Let $\widehat{F}_{H}$ be a smooth function on $[0,+\infty)$ such that $\widehat{F}_{H}(E)\geq F_{H}(E)$ for all $E\geq E_0$  possessing the properties
\begin{equation}\label{F-prop-1}
\widehat{F}_{H}(E)>0,\quad \widehat{F}_{H}^{\shs\prime}(E)>0,\quad \widehat{F}_{H}^{\shs\prime\prime}(E)\leq 0\quad\textrm{for all }\; E>0.
\end{equation}
At least one such function $\widehat{F}_{H}$ always exists: the  function  $E\mapsto F_{H}(E+E_0)$ satisfies all the above conditions by Proposition 1 in \cite{EC}.

The metric divergence of a generalized ensemble $\mu$ with respect to a state $\sigma$ is naturally defined as
\begin{equation}\label{r-amd+}
  D(\mu|\shs\sigma)=\textstyle\frac{1}{2}\displaystyle\int\|\rho-\sigma\|_1\mu(d\rho).
\end{equation}
If $\mu=\{p_i,\rho_i\}$ then (\ref{r-amd+}) coincides with (\ref{r-amd}). \smallskip

\begin{property}\label{hub-ce} \emph{Let $\,\mu\,$ be a generalized ensembles of states in $\,\S(\H)$ with finite average energy $\bar{E}(\mu)\doteq E(\bar{\rho}(\mu))$ and  $\,\sigma$ a state in $\,\S(\H)$ with finite energy $E(\sigma)$. Let  $\,\varepsilon=D(\mu|\shs\sigma)$ be the metric divergence of $\,\mu$ with respect to $\sigma$ defined in (\ref{r-amd+}). Then
\begin{equation}\label{hub-ce+}
\!\chi(\mu)\leq \min_{t\in(0,a]}\left(\varepsilon\!\left(\frac{1+\kappa t}{1-\varepsilon t}+t\right)\!
\widehat{F}_{H}\!\left(\frac{\bar{E}(\mu)}{\varepsilon t}\right)+
h_2(\varepsilon t)+g\!\left(\frac{1+\kappa t}{1-\varepsilon t}\,\varepsilon\!\right)\!\right)\!,
\end{equation}
where  $\,a=1/(2\varepsilon)$, $\widehat{F}_{H}$ is any upper bound for the function $F_{H}$ (defined in (\ref{F-def})) satisfying conditions (\ref{F-prop-1}) and $\,\kappa=\frac{1}{2}(1+E(\sigma)/\bar{E}(\mu))$.}\footnote{$h_2(p)$ is the binary entropy, $g(p)=(1+p)h_2\!\left(\frac{p}{1+p}\right)=(p+1)\log(p+1)-p\log p$.}
\end{property}\medskip

\begin{remark}\label{os}
The right hand side of (\ref{hub-ce+}) is an increasing function of $\varepsilon$. It tends to zero as
$\,\varepsilon\rightarrow0\,$ if and only if $\,\widehat{F}_{H}(E)=o\shs(E)$ as $E\rightarrow+\infty$. By Proposition 1 in \cite{EC} the function
$\,\widehat{F}_{H}(E)=F_{H}(E+E_0)$ satisfies the last condition if and only if
\begin{equation}\label{g-cond+}
\Tr e^{-\lambda H}<+\infty\,\textrm{ for all }\,\lambda>0.
\end{equation}
It is interesting that (\ref{g-cond+}) is a necessary and sufficient condition of continuity of the Holevo quantity on the set of all generalized ensembles $\mu$ with bounded average energy $\bar{E}(\mu)$ with respect to the weak convergence topology. This follows from Proposition 8 in \cite{CHI}, since (\ref{g-cond+}) is  a necessary and sufficient condition of continuity of the von Neumann entropy on the set of states $\rho$ with bounded energy $E(\rho)=\Tr H\rho$ \cite{W,EC}.
\end{remark}\smallskip

\emph{Proof.} Assume first that $\mu$ is a discrete ensemble $\{p_i,\rho_i\}$ with the average state $\bar{\rho}$.

Following the proofs of Lemmas 16,17 in \cite{W-CB} take any $\delta\in(0,\frac{1}{2}]$ and denote by $P_{\delta}$  the spectral projector of the operator $H$ corresponding to the interval $[0, \delta^{-1}\bar{E}(\mu)]$. By condition (\ref{g-cond}) $\Tr P_{\delta}<+\infty$.
Since $\Tr H\bar{\rho}=\bar{E}(\mu)$ and $\Tr H\sigma=E(\sigma)$, it is easy to show that
\begin{equation}\label{ce-t-1}
\Tr P_{\delta}\bar{\rho}\geq 1-\delta \quad \textrm{and} \quad \Tr P_{\delta}\sigma\geq 1-\delta E(\sigma)/\bar{E}(\mu).
\end{equation}

Consider the ensemble $\{\hat{p}_i,\hat{\rho}_i\}$, where $\hat{\rho}_i=r_i^{-1}P_{\delta}\rho_iP_{\delta}$, $\hat{p}_i=r_i p_i/r$, $r_i=\Tr P_{\delta}\rho_i$, $r=\Tr P_{\delta}\bar{\rho}$.  Corollary \ref{amd-ub} implies
\begin{equation}\label{ce-t-2}
 \chi(\{\hat{p}_i,\hat{\rho}_i\})\leq \hat{\varepsilon} \log\Tr P_{\delta}+g(\hat{\varepsilon}),
\end{equation}
where $\hat{\varepsilon}$ is the average metric divergence of the ensemble $\{\hat{p}_i,\hat{\rho}_i\}$.

Let $\hat{\sigma}=s^{-1}P_{\delta}\sigma P_{\delta}$, where $s=\Tr P_{\delta}\sigma$. Then
\begin{equation}\label{ce-t-2+}
\begin{array}{rl}
2\hat{\varepsilon}\,\leq & \!\!\sum_i \hat{p}_i\|\hat{\rho}_i- \hat{\sigma}\|_1=r^{-1}\sum_i p_i\|P_{\delta}\rho_iP_{\delta}-(r_i/s) P_{\delta}\sigma P_{\delta}\|_1\\\\
\leq &\!\! r^{-1}\sum_i p_i\!\left(\|P_{\delta}\rho_iP_{\delta}-P_{\delta}\sigma P_{\delta}\|_1+|1-(r_i/s)|\| P_{\delta}\sigma P_{\delta}\|_1\right)\\\\
\leq & \!\! r^{-1}\sum_i p_i\!\left(\|\rho_i-\sigma\|_1+|(1-r_i)-(1-s)|\right)\\\\ \leq & \!\! r^{-1}(2\varepsilon+(1-\Tr P_{\delta}\bar{\rho})+(1-\Tr P_{\delta}\sigma))
\leq  (1-\delta)^{-1}(2\varepsilon+2\kappa\delta),
\end{array}
\end{equation}
where the last inequality  follows from (\ref{ce-t-1}).

By using  (\ref{ce-t-1}) and the arguments from the proof of Lemma 16 in \cite{W-CB} (based on properties (\ref{F-prop-1}) of the function $\widehat{F}_{H}$) we obtain
\begin{equation*}
  H(\bar{\rho})-H(P_{\delta}\bar{\rho}P_{\delta})\leq \delta \widehat{F}_{H}(\bar{E}(\mu)/\delta)+h_2(\delta).
\end{equation*}
This inequality and Lemma 2 in \cite{CHI} imply that
\begin{equation}\label{ce-t-3}
\chi(\{p_i,\rho_i\})-\chi(\{\hat{p}_i,\hat{\rho}_i\})\leq\delta \widehat{F}_{H}(\bar{E}(\mu)/\delta)+h_2(\delta).
\end{equation}

Since the energy of the state $[\Tr P_{\delta}]^{-1}P_{\delta}$ does not exceed $\bar{E}(\mu)/\delta$, its entropy
$\log\Tr P_{\delta}$ is upper bounded by $\widehat{F}_{H}(\bar{E}(\mu)/\delta)$. So,
it follows from (\ref{ce-t-2}), (\ref{ce-t-2+}) and (\ref{ce-t-3}) that
\begin{equation}\label{hub-ce-}
\!\chi(\{p_i,\rho_i\})\leq(\varepsilon'+\delta)\widehat{F}_{H}\!\left(\frac{\bar{E}(\mu)}{\delta}\right)+g(\varepsilon')+h_2(\delta),\;\;\textrm{where}\;\; \varepsilon'=\frac{\varepsilon+\kappa\delta}{1-\delta}.
\end{equation}

Now assume that $\delta=\varepsilon t$, where $t\in(0,\frac{1}{2\varepsilon}]$. Then $\varepsilon'=\varepsilon (1+\kappa t)/(1-\varepsilon t)$
and hence (\ref{hub-ce-}) implies (\ref{hub-ce+}) for $\mu=\{p_i,\rho_i\}$. \smallskip

For arbitrary generalized ensemble $\mu$ there exists a sequence $\{\mu_n\}$ of discrete ensembles weakly\footnote{The weak convergence of a sequence $\{\mu_n\}$  to an ensemble $\mu_0$ means that
$\,\lim_{n\rightarrow\infty}\int f(\rho)\mu_n(d\rho)=\int f(\rho)\mu_0(d\rho)\,$
for any continuous bounded function $f$ on $\,\S(\H)$.} converging to $\mu$ such that
$$
\lim_{n\rightarrow\infty}\chi(\mu_n)=\chi(\mu)\quad\textrm{ and }\quad \bar{\rho}(\mu_n)=\bar{\rho}(\mu)\,\textrm{ for all }n.
$$
Such sequence can be obtained by using the construction from the proof of Lemma 1 in \cite{H-Sh-2} and taking into account the lower semicontinuity of the function $\mu\mapsto\chi(\mu)$ \cite[Pr.1]{H-Sh-2}. Since $D(\mu_n|\shs\sigma)$ tends to $D(\mu|\shs\sigma)$ (due to the weak convergence of $\mu_n$ to $\mu$), the validity of inequality (\ref{hub-ce+}) for the ensemble $\mu$ follows from its validity for all the ensembles $\mu_n$ proved before. $\square$
\medskip

Consider specification of the upper bound in Proposition \ref{hub-ce}  for the $\,\ell$-mode quantum oscillator. In this case
\begin{equation}\label{osc-H}
H=\sum_{i=1}^{\ell}\hbar\shs\omega_i \left (a^{+}_ia_i+\textstyle\frac{1}{2}I_A\right),\quad E_0\doteq\frac{1}{2}\sum_{i=1}^{\ell}\hbar\omega_i,
\end{equation}
where $\,a_i\,$ and $\,a^{+}_i\,$ are the annihilation and creation operators and $\,\omega_i\,$ is the frequency of the $i$-th oscillator \cite[Ch.12]{H-SCI}. Since condition (\ref{g-cond+}) holds, for any $E>E_0$ the von Neumann entropy $H(\rho)$ is continuous on the sets of states determined by the inequality $\Tr H\rho\leq E$ and attains maximum on this set at the Gibbs state $\gamma(E)=[\Tr e^{-\lambda(E)H}]^{-1}e^{-\lambda(E)H}$, where $\lambda(E)$ is the solution of the equation $\Tr He^{-\lambda H}=E\Tr e^{-\lambda H}$ \cite{W}. \smallskip

The exact value of $F_{H}(E)\doteq\sup_{\Tr H\rho\leq E}H(\rho)$ can be found by solving
a transcendental equation. But one can show that $F_{H}(E)$ is upper bounded by the function
\begin{equation}\label{bF-ub}
\widehat{F}_{\ell,\omega}(E)\doteq\ell\log \frac{E+E_0}{\ell E_*}+\ell,\quad E_*=\left[\prod_{i=1}^{\ell}\hbar\omega_i\right]^{1/\ell},
\end{equation}
on $[0,+\infty)$ satisfying conditions (\ref{F-prop-1}) such that
$\,\widehat{F}_{\ell,\omega}(E)-F_{H}(E)$ tends to zero as $\,E\rightarrow+\infty$ \cite[Sect.3.2]{CHI}.
\smallskip

\begin{corollary}\label{hub-ce++}
\emph{Let $\,\mu\,$ be a generalized ensembles of states of the $\,\ell$-mode quantum oscillator with finite average energy $\bar{E}(\mu)\doteq E(\bar{\rho}(\mu))$ and  $\,\sigma$ a state with finite energy $E(\sigma)$. Let  $\,\varepsilon=D(\mu|\shs\sigma)$ be the metric divergence of $\,\mu$ with respect to $\,\sigma$ defined in (\ref{r-amd+}). Then
\begin{equation*}
\!\chi(\mu)\leq \min_{t\in(0,a]}\!\left(\varepsilon\!\left(\frac{1+\kappa t}{1-\varepsilon t}+t\right)\!\left[\widehat{F}_{\ell,\omega}(\bar{E}(\mu))-\ell\log(\varepsilon t)\right]
+h_2(\varepsilon t)+g\!\left(\frac{1+\kappa t}{1-\varepsilon t}\,\varepsilon\!\right)\!\right),
\end{equation*}
where  $\,a=1/(2\varepsilon)$, $\widehat{F}_{\ell,\omega}(E)$ is defined in (\ref{bF-ub}) and $\,\kappa=\frac{1}{2}(1+E(\sigma)/\bar{E}(\mu))$.}\smallskip

\emph{This upper bound is tight (for large $E$ and appropriate  choice of $\,\sigma$).}
\end{corollary}\medskip

\emph{Proof.} Since
$\widehat{F}_{\ell,\omega}(E/x)\leq \widehat{F}_{\ell,\omega}(E)-\ell\log x$ for any positive $E$ and $x\leq1$,
the main assertion of the corollary directly follows from Proposition \ref{hub-ce}.

Let $E>E_0$ and $\{p_i,\rho_i\}$ be any pure state ensemble with the average state $\gamma(E)$.
Consider the ensemble $\{p_i,\rho^{\varepsilon}_i\}$, where $\rho^{\varepsilon}_i=\varepsilon\rho_i+(1-\varepsilon)\gamma(E)$.
Then
$$
2D(\{p_i,\rho^{\varepsilon}_i\}|\shs\gamma(E))=\sum_{i}p_i\|\shs \rho^{\varepsilon}_i-\gamma(E)\|_1=\sum_{i}\varepsilon p_i\|\shs\rho_i-\gamma(E)\|_1\leq2\varepsilon,
$$
while concavity of the entropy implies
\begin{equation}\label{tmp}
\chi(\{p_i,\rho^{\varepsilon}_i\})\geq \varepsilon H\!\left(\gamma(E)\right)-h_2(\varepsilon)=\varepsilon F_{H}(E)-h_2(\varepsilon).
\end{equation}
This shows  tightness of the upper bound, since  $\,\widehat{F}_{\ell,\omega}(E)-F_{H}(E)=o(1)\,$ as $\,E\rightarrow+\infty$ and
the quantity
$$
\varepsilon\!\left(\frac{1+ t}{1-t}+t\right)\!\left[\widehat{F}_{\ell,\omega}(E)-\ell\log(\varepsilon t)\right]
$$
can be made not greater than $\,\varepsilon (\widehat{F}_{\ell,\omega}(E)+o\shs(\widehat{F}_{\ell,\omega}(E)))\,$  as $E\rightarrow+\infty$ by appropriate choice of $\,t$. This follows from Lemma \ref{s-l} below proved by elementary methods. $\square$

\begin{lemma}\label{s-l} \emph{Let $\,f(t)=\frac{1+t}{1-t}+t$, $\,b>0\,$ and $\,c\,$ be arbitrary. Then
$$
\min_{t\in(0,\frac{1}{2})}f(t)(x-b\log t+c)\leq x+o(x)\quad\textrm{as}\quad x\rightarrow+\infty.
$$}
\end{lemma}

\section{Upper bounds for the Holevo capacity}

\subsection{Finite-dimensional channels}

A \emph{quantum channel} $\,\Phi$ from a system $A$ to a system
$B$ is a completely positive trace preserving linear map
$\mathfrak{T}(\mathcal{H}_A)\rightarrow\mathfrak{T}(\mathcal{H}_B)$,
where $\mathcal{H}_A$ and $\mathcal{H}_B$ are Hilbert spaces
associated with these systems \cite{H-SCI,N&Ch,Wilde}.\smallskip

The \emph{Holevo capacity} of a quantum channel
$\Phi:A\rightarrow B$  is
defined as follows
\begin{equation}\label{HC-def}
C_{\chi}(\Phi)=\sup_{\{p_i,\rho_i\}}\chi(\{p_i,\Phi(\rho_i)\}),
\end{equation}
where the supremum is over all
ensembles of input states. This quantity determines the ultimate rate of transmission of classical information trough the channel $\Phi$ with non-entangled input encoding, it is closely related to the classical capacity of a quantum channel \cite{H-SCI,N&Ch,Wilde}.\smallskip

For a given subset $\S_0$ of $\S(\H)$ consider the quantity
$$
Cr(\S_0)\doteq \textstyle\frac{1}{2}\displaystyle\inf_{\sigma\in\S(\H)}\sup_{\rho\in\S_0}\|\rho-\sigma\|_1
$$
called Chebyshev radius of $\S_0$ with respect to the metric $\Delta(\rho,\sigma)=\frac{1}{2}\|\rho-\sigma\|_1$  \cite{cr-1,cr-2}. For example,
$Cr(\{\rho,\sigma\})=\frac{1}{4}\|\rho-\sigma\|_1$ and $Cr(\S(\H))=1-1/d$, where $d=\dim\H$. The Chebyshev radius of a set $\S_0$ does not exceed its diameter $D(\S_0)\doteq \textstyle\frac{1}{2}\displaystyle\sup_{\rho,\sigma\in\S_0}\|\rho-\sigma\|_1$, but $Cr(\S_0)$ may be significantly less than $D(\S_0)$ even for multi-dimensional sets $\S_0$: the diameter of the set of vectors in Example \ref{MMD-e} is equal to $a\sqrt{2-a^2}$ while its Chebyshev radius is less than $a$.\smallskip

Corollary \ref{amd-ub} implies the following\smallskip
\begin{property}\label{chi-cap} \emph{Let $\,\Phi:A\rightarrow B\,$ be a quantum channel. Then
\begin{equation}\label{chi-cap-ub}
C_{\chi}(\Phi)\leq r_{\Phi}\log d_B+g(r_{\Phi}),
\end{equation}
where $\,r_{\Phi}=Cr(\Phi(\S(\H_A)))\,$ and  $\,d_B=\dim\H_B$.\footnote{$g(p)=(1+p)h_2\!\left(\frac{p}{1+p}\right)=(p+1)\log(p+1)-p\log p$.} Upper bound (\ref{chi-cap-ub}) is tight.}\smallskip
\end{property}

\emph{Proof.} Inequality (\ref{chi-cap-ub}) follows from the second inequality in Corollary \ref{amd-ub}, since the average metric divergence $\varepsilon_{\mathrm{av}}$ of the image of any input ensemble $\,\{p_i,\rho_i\}$ under the channel $\Phi$ does not exceed $r_{\Phi}$.

The tightness of upper bound (\ref{chi-cap-ub}) follows from Examples \ref{chi-e1} and \ref{chi-e2} below. $\square$ \smallskip

\begin{remark}\label{chi-r}
By Corollary \ref{amd-ub} the quantity $\,r_{\Phi}=Cr(\Phi(\S(\H_A)))\,$ in  (\ref{chi-cap-ub}) can be replaced by the quantity
$\frac{1}{2}\sup_{\{p_i,\rho_i\}}\inf_{\sigma\in\S(\H_B)}\sum_ip_i\|\Phi(\rho_i)-\sigma\|_1$ which  formally may be  less than $r_{\Phi}$. But we have not found examples for which this quantity is really less than $r_{\Phi}$.\smallskip
\end{remark}

The following example shows that the extra term $g(r_{\Phi})$ in (\ref{chi-cap-ub}) can not be removed. \smallskip
\begin{example}\label{chi-e1}
Let $\Phi:A\rightarrow B$ be a quantum channel such that the set $\Phi(\S(\H_A))$ contains a collection of pure states corresponding to some orthonormal basis in $\H_B$ (for example, $\Phi$ is the identity channel or the channel $\rho\mapsto\sum_{k}\langle \varphi_k|\rho|\varphi_k\rangle|\psi_k\rangle\langle\psi_k|$, where $\{|\varphi_k\rangle\}$ and $\{|\psi_k\rangle\}$ are orthonormal base in $\H_A$ and $\H_B\cong\H_A$ correspondingly). Then $C_{\chi}(\Phi)=\log d_B$ and $r_{\Phi}=1-1/d_B$. So, in this case inequality (\ref{chi-cap-ub}) has the form
$$
C_{\chi}(\Phi)=\log d_B\leq (1-1/d_B)\log d_B+g(1-1/d_B),
$$
which would not be valid without the term $g(1-1/d_B)$. $\square$
\end{example}\smallskip

Despite the fact that upper bound (\ref{chi-cap-ub}) depends only on the Chebyshev radius of the output set of a channel $\Phi$, it gives
relatively sharp estimates for the Holevo capacity of some nontrivial channels. \smallskip

\begin{example}\label{chi-e2}
Let $\Phi_p$ be a depolarizing channel from $d$-dimensional quantum system to itself, i.e.
$\Phi_p(\rho)=(1-p)\rho+p\rho_c$, where $\rho_c$ is the chaotic state and $p\in[0,1]$. Then
$$
C_{\chi}(\Phi_p)=(1-pc)\log d-h_2(pc)-pc\log c,
$$
where $\,c=1-1/d$ \cite{H-SCI,Wilde}, while the upper bound (\ref{chi-cap-ub}) implies
$$
C_{\chi}(\Phi_p)\leq (1-pc)\log d+g((1-p)c)-(1/d)\log d ,
$$
since $\|\Phi_p(\rho)-\rho_c\|_1=(1-p)\|\rho-\rho_c\|_1\leq (1-p)c\,$ for any input state $\rho$.
\end{example} \medskip

Another example for which upper bound (\ref{chi-cap-ub}) gives asymptotically sharp estimates for the Holevo capacity is the erasure channel
\begin{equation*}
\Psi_p(\rho)=\left[\begin{array}{cc}
(1-p)\rho &  0 \\
0 &  p\Tr\rho
\end{array}\right], \quad p\in[0,1],
\end{equation*}
from $d$-dimensional quantum system to its $(d+1)$-dimensional extension, since in this case
$\,C_{\chi}(\Psi_p)=(1-p)\log d\,$ and $\,r_{\Psi_p}=(1-p)(1-1/d)$.\medskip

The following example shows that accuracy of the upper bound (\ref{chi-cap-ub}) varies significantly within one class of channels.\smallskip

\begin{example}\label{chi-e3}
Let $\Phi:A\rightarrow B$ be a quantum channel such that the set $\Phi(\S(\H_A))$ coincides with the convex hull of a set $\S_0$ of isomorphic states in $\S(\H_B)$ and contains the chaotic state $\rho_c\doteq I_{\H_B}/d_B$, where $d_B=\dim\H_B$ (for example, $\Phi$ is the channel $\rho\mapsto\sum_{k}\langle \varphi_k|\rho|\varphi_k\rangle \sigma_k$, where $\{|\varphi_k\rangle\}$ is an orthonormal basis in $\H_A$ and $\{\sigma_k\}$ is a collection of isomorphic states in $\S(\H_B)$ such that $\,\rho_c=\sum_kp_k\sigma_k\,$ for some probability distribution $\{p_k\}$).
Then $C_{\chi}(\Phi)=\log d_B-H_{\mathrm{min}}(\Phi)$, where $H_{\mathrm{min}}(\Phi)=H(\sigma),\; \sigma\in\S_0$.\smallskip

We will show that accuracy of the upper bound (\ref{chi-cap-ub}) strongly depends on the form of spectrum of the states in $\S_0$.\smallskip

Assume first that all the states in $\S_0$ have the spectrum
$$
\{\,1-r/d,\, \underbrace{1/d,...,1/d}_{r},\, \underbrace{0,...,0}_{d-r-1}\,\},
$$
where $d=d_B$ and $r<d-1$. In this case $H_{\mathrm{min}}(\Phi)=(r/d)\log d+\eta(1-r/d)$  and
hence
$$
C_{\chi}(\Phi)=(1-r/d)\log d-\eta(1-r/d),
$$
while upper bound (\ref{chi-cap-ub}) implies
$$
C_{\chi}(\Phi)\leq(1-r/d-1/d)\log d+g(1-r/d-1/d),
$$
since $\,\|\sigma-\rho_c\|_1=2(d-r-1)/d\,$ for all $\sigma\in\S_0$. We see again that upper bound (\ref{chi-cap-ub}) gives asymptotically sharp estimate for the Holevo capacity for large $d$ and any $r$.\smallskip

Now assume that all the states in $\S_0$ are proportional to $r$-rank projectors. Then
$$
C_{\chi}(\Phi)=\log d-\log r,
$$
while the upper bound (\ref{chi-cap-ub}) implies
$$
C_{\chi}(\Phi)\leq(1-r/d)\log d+g(1-r/d).
$$
So, in this case the upper bound (\ref{chi-cap-ub}) gives too rough estimate for the Holevo capacity.
\end{example}\smallskip

\subsection{Infinite-dimensional channels with energy constraints}

The Holevo capacity of an infinite-dimensional quantum channel $\Phi:A\rightarrow B$ with energy constraint can be defined as follows
\begin{equation}
C_{\chi}(\Phi,H_A,E)=\sup_{\bar{E}(\mu)\leq E}\chi(\Phi(\mu)),
\label{chi-ec-cap-d}
\end{equation}
where $H_A$ is the Hamiltonian of the system $A$, the supremum is over all generalized input ensembles $\mu$ with the average energy $\bar{E}(\mu)\doteq\mathrm{Tr}H_A\bar{\rho}(\mu)$ not exceeding $E$ and $\Phi(\mu)$ is the image of $\mu$ under the channel $\Phi$ (defined as the measure $\mu\circ\Phi^{-1}$ on $\S(\H_B)$). In fact, the supremum in (\ref{chi-ec-cap-d}) can be taken only over discrete ensembles \cite{H-Sh-2}. This quantity determines the ultimate rate of transmission of classical information trough the channel $\Phi$ under the constraint
on mean energy of a code if only  non-entangled input encoding is used \cite[Ch.12]{H-SCI}.

For given channel $\Phi:A\rightarrow B$ and state $\sigma$ in $\,\S(\H_B)$ introduce the quantity
\begin{equation}\label{amd-ch}
  D(\Phi|\shs\sigma)=\textstyle\frac{1}{2}\displaystyle\sup_{\rho\in\S(\H_A)}\|\Phi(\rho)-\sigma\|_1,
\end{equation}
which can be called output metric divergence of $\Phi$ with respect to $\sigma$.\smallskip

Assume that the Hamiltonian $H_B$ of the system $B$ satisfies condition
(\ref{g-cond}). Denote by $E_X(\rho)$ the energy $\Tr H_X \rho$ of a state $\rho$ in $\S(\H_X)$, $X=A,B$.\smallskip

\begin{property}\label{chi-ec-cap-ub} \emph{Let $\Phi:A\rightarrow B$ be a quantum channel and  $\sigma$ a state in $\,\S(\H_B)$ with finite energy $E_B(\sigma)$. Let  $\,\varepsilon=D(\Phi|\shs\sigma)$ be the output metric divergence of $\,\Phi$ with respect to $\sigma$ defined in (\ref{amd-ch}). If $\,E_*=\sup\limits_{E_A(\rho)\leq E} E_B(\Phi(\rho))\,$ is finite then
\begin{equation*}
C_{\chi}(\Phi,H_A,E)\leq \min_{t\in(0,a]}\left(\varepsilon\!\left(\frac{1+\kappa t}{1-\varepsilon t}+t\right)\!
\widehat{F}_{H_B}\!\left(\frac{E_*}{\varepsilon t}\right)+
h_2(\varepsilon t)+g\!\left(\frac{1+\kappa t}{1-\varepsilon t}\,\varepsilon\!\right)\!\right),\!
\end{equation*}
where  $\,a=1/(2\varepsilon)$, $\widehat{F}_{H_B}$ is any upper bound for the function $F_{H_B}$ (defined in (\ref{F-def})) satisfying conditions (\ref{F-prop-1}) and $\,\kappa=\frac{1}{2}(1+E_B(\sigma)/E_*)$.}\footnote{$h_2(p)$ is the binary entropy, $g(p)=(1+p)h_2\!\left(\frac{p}{1+p}\right)=(p+1)\log(p+1)-p\log p$.}\smallskip

\emph{If $\,B$ is the $\,\ell$-mode quantum oscillator and $\,\widehat{F}_{H_B}=\widehat{F}_{\ell,\omega}$\footnote{The function $\widehat{F}_{\ell,\omega}$ is defined in (\ref{bF-ub}).} then the above upper bound for  $C_{\chi}(\Phi,H_A,E)$ is tight (for large $E$ and optimal  choice of $\,\sigma$).}
\end{property}\medskip

\emph{Proof.} The main assertion of the proposition directly follows from Proposition \ref{hub-ce} and definition (\ref{chi-ec-cap-d}) of the Holevo capacity.

The last assertion follows from Example \ref{last-ex} below. $\square$\smallskip

\begin{example}\label{last-ex} Let $A=B$ be the $\,\ell$-mode quantum oscillator. Consider the channel
$$
\Phi^\sigma_p(\rho)=(1-p)\rho+p\shs\sigma,
$$
where $\sigma$ is a given state with finite energy $E(\sigma)$ and $p\in[0,1]$.

By using joint convexity of the relative entropy, concavity of the von Neumann entropy and inequality (\ref{H-ineq}) one can show that
\begin{equation}\label{d-ineq}
(1-p)H(\gamma_A(E))-h_2(p)\leq C_{\chi}(\Phi^\sigma_p,H_A,E)\leq (1-p)H(\gamma_A(E)),
\end{equation}
where $\gamma_A(E)$ is the Gibbs states of the system $A=B$ corresponding to the energy $E$.

In this case $D(\Phi^\sigma_p|\shs\sigma)\leq (1-p)$ and $E_*=(1-p)E+pE(\sigma)$. Assume for simplicity that $E(\sigma)\leq E$. Then
$E_*\leq E$ and Proposition \ref{chi-ec-cap-ub} with
$\,\widehat{F}_{H_B}=\widehat{F}_{\ell,\omega}$ gives the
upper bound
\begin{equation}\label{chi-ec-cap-ub+}
\!\begin{array}{ccc}
\displaystyle C_{\chi}(\Phi^\sigma_p,H_A,E)\leq\min_{t\in(0,a]}\left(\bar{p}\shs(f_p(t)+t)\widehat{F}_{\ell,\omega}\!\left(\frac{E}{t\bar{p}}\right)+
h_2(t\bar{p})+g(f_p(t)\bar{p})\right)\\\\ \displaystyle \leq\min_{t\in(0,a]}\left(\bar{p}\shs(f_p(t)+t)\!\left[\widehat{F}_{\ell,\omega}(E)-\ell\log(t\bar{p})\right]+
h_2(t\bar{p})+g(f_p(t)\bar{p})\right),
\end{array}\!\!
\end{equation}
where $\,a=\frac{1}{2(1-p)}$, $\,\bar{p}=1-p$ and $\,f_p(t)=(1+t)/(1-(1-p)t)$. By Lemma \ref{s-l} the right hand side of (\ref{chi-ec-cap-ub+}) is equal to
\begin{equation*}
(1-p)\widehat{F}_{\ell,\omega}(E)+o\shs(\widehat{F}_{\ell,\omega}(E))\quad \textrm{as}\;\; E\rightarrow+\infty.
\end{equation*}
Since $\widehat{F}_{\ell,\omega}(E)-H(\gamma_A(E))=o(1)$ as $E\rightarrow+\infty$, comparing this  with (\ref{d-ineq}) we see that the upper bound (\ref{chi-ec-cap-ub+}) is tight for large $E$. $\square$
\end{example}

\bigskip
I am grateful to A.S.Holevo and G.G.Amosov for useful discussion.

\end{document}